# Halting hyaluronidase activity with hyaluronan-based nanohydrogels: development of versatile injectable formulations

E. Montanari[a], N. Zoratto[a], L. Mosca[b], L. Cervoni[b], E. Lallana[c], R. Angelini[d,e], R. Matassa[f], T. Coviello[a], C. Di Meo[a]*, P. Matricardi[a],

[a]Department of Drug Chemistry and Technologies, Sapienza University of Rome, P.le Aldo Moro 5, Rome 00185, Italy

[b]Department of Biochemical Sciences, Sapienza University of Rome, P.le Aldo Moro 5, Rome 00185, Italy

[c]Faculty of Biology, Medicine and Health, The University of Manchester, Oxford road, M13 9PT, Manchester, UK

[d]Istituto dei Sistemi Complessi del Consiglio Nazionale delle Ricerche (ISC-CNR), P.le Aldo Moro 5, Rome I-00185, Italy

[e]Department of Physics, Sapienza University of Rome, P.le Aldo Moro 5, Rome 00185, Italy

[f]Department of Anatomical, Histological, Forensic and Orthopaedic Sciences, Sapienza University of Rome, Via A. Borelli, Rome 00161, Italy.

*Corresponding author. Tel: +39 0649913226, E-mail address: chiara.dimeo@uniroma1.it

ABSTRACT

Hyaluronan (HA) is among the most used biopolymers for viscosupplementation and dermo-cosmetic applications. However, the current injectable HA-based formulations present relevant limitations: I) unmodified HA is quickly degraded by endogenous hyaluronidases (HAase), resulting in short lasting properties; II) cross-linked HA, although shows enhanced stability against HAase, often contains toxic chemical cross-linkers. As such, herein, we present biocompatible self-assembled hyaluronan-cholesterol nanohydrogels (HA-CH NHs) able to bind to HAase and inhibit the enzyme activity *in vitro*, more efficiently than currently marketed HA-based cross-linked formulations (e.g. Jonexa[TM]). HA-CH NHs inhibit HAase through a mixed mechanism, by which NHs bind to HAase with an affinity constant 7-fold higher than that of native HA. Similar NHs, based on gellan-cholesterol, evidenced no binding to HAase, neither inhibition of the enzyme activity, suggesting this effect might be due to the specific binding of HA-CH to the active site of the enzyme. Therefore, HA-CH NHs were engineered into injectable hybrid HA mixtures or physical hydrogels, able to halt the enzymatic degradation of HA.

**Abbreviations**

Hyaluronan (HA); hyaluronidase (HAase); hyaluronan-cholesterol (HA-CH); gellan-cholesterol (Ge-CH); nanohydrogels (NHs); physical hydrogels (PHs); hybrid hyaluronan mixtures (HHs); Jonexa$^{TM}$ Hylastan SGL-80 (Hylastan); osteoarthritis (OA).

1. Introduction

The global market of hyaluronan (HA) was worth 7.2 billion USD in 2016 and it is expected to increase up to 15.4 USD billion by 2025. This continuous growth is essentially due to the rising number of people affected by osteoarthritis (OA)[1, 2] and also to the increasing aesthetic procedures using HA-based materials (e.g. dermal fillers[3]). In 1997, the first injectable form of HA (Hyalgan®)[4] was approved by the Food and Drug Administration (FDA) for the treatment of knee OA, which is a disease causing the degradation of joints, intense pain and eventually loss of function[5]. OA is usually treated with both corticosteroid (e.g. betamethasone) and HA injections into the synovial joint[5]; the latter mimics the natural presence of HA in the synovial fluid[6] and prevents further erosion of the joint by increasing its lubrication that is severely depleted in patients suffering from OA. On the other hand, Hylaform® was the first HA-based dermal filler approved in 2006 by the United States[7]. A number of features make HA an attractive biomaterial in medicine and cosmetics: its high hydrophilicity, its natural presence in the body (e.g. skin and synovial joints) and its low propensity to elicit adverse reactions (e.g. immunologic reactions). Despite these advantages, HA also presents several limitations, most importantly, free HA (i.e. the uncross-linked and unmodified polymer) is rapidly degraded through different pathways[8] (e.g. *via* endogenous hyaluronidase (HAase) enzymes and/or *via* free radicals) and, thus, it is eliminated from the injection site in less than one week[9]. To overcome the lack of persistence of HA solutions in the body, viscosupplements and dermal fillers were prepared containing HA chains cross-linked with specific cross-linking agents[3]. The obtained hydrogels ensure a less enzymatic degradation within the injection site, persisting for longer period of times. Currently, two chemical cross-linkers dominate the US market: 1,4-butanediol diglycidyl ether (BDDE)[10] and divinyl sulfone (DVS)[11]. Both of these chemicals react with the HA hydroxyl groups, leading to the formation of three-dimensional networks, which are able to slow down the degradation and the consequent drainage of HA from the skin and synovial joints. A number of cross-linked HA-based materials, with different degrees of cross-linking, different molecular weights (MW) and different concentrations are currently in the market[9]; typically, they are hybrid materials composed of both HA cross-linked and uncross-linked fractions (usually, less than 20% (wt.%) of HA molecules are cross-linked and suspended in a HA solution). Nevertheless, the use of chemical cross-linkers suffers from several drawbacks: I) unreacted or residual cross-linker molecules are toxic and may trigger side effects; II) purification steps usually result in relatively low efficiency and high costs; III) cross-linking inevitably increases HA viscosity, resulting in painful injections and lowering patient compliance, thus limiting the maximum tolerable concentration of the injectable material[3]. To overcome the above drawbacks, herein we describe the ability of self-assembled hyaluronan-cholesterol (HA-CH) nanohydrogels (NHs) to bind to HAase and to halt its enzymatic activity *in vitro*. In this respect, the attachment of cholesterol (CH) to the HA chains allows the polymer to assemble spontaneously in aqueous environment forming NHs, which are well-defined spherical nano-sized tri-dimensional networks, composed by internal hydrophobic domains (e.g., cholesteryl groups) and an outer hydrophilic shell (e.g., HA chains), able to swell in aqueous environment, to absorb a large amount of water[12, 13] and to entrap a wide range of active compounds, such as hydrophobic[14] or hydrophilic[15] drugs, as well as (poly)peptides[16]. Among the hydrophobic moieties, CH is a biocompatible and freely available compound, which naturally occurs in skin and cell membranes[17]. Together with ceramide and fatty acids, CH plays a key role in maintaining the normal hydration levels required for the physiological functions of the skin[18]; as such, CH is usually used as skin-conditioning agent in cosmetics. Herein, HA-CH NHs were mixed with a HA solution, forming hybrid hyaluronan mixtures (HHs) or were re-suspended at concentrations higher than 6% (w/v), forming physical hydrogels (PHs) in physiological medium. These two systems showed a number of advantages compared to the commercial cross-linked HA (e.g., Jonexa™ Hylastan SGL-80 (Hylastan)), such as: I) the chance to avoid the use of chemical cross-linkers; II) the possibility to increase the HA concentration without affecting the HA solution viscosity; III) more importantly, the ability to halt the HAase activity *in vitro*, leading to the development of more efficient HAase-resistant materials. HAases are the main responsible for the enzymatic degradation of HA in the body, although they have also a limited ability to degrade chondroitin and chondroitin sulphate[19]. Human HAases are a group of five endo- -N-acetyl-hexosaminidases (HAase-1, -2, -3, -4, and PH-20), which degrade HA by using a hydrolytic mechanism of action not yet fully understood[20,21]. HAase-1 and 2 are the two major HAases present in human tissues. HAase-2 degrades high MW HA to polymers with MW of approximately $2.0 \times 10^3$ (~ 50 HA repeating units), whereas HAase-1 can degrade high MW HA to mainly tetrasaccharides. Interestingly, the HAase active site is lined by a number of cationic and hydrophobic amino acids, which ideally should allow the binding to anionic and hydrophobic substrates.

**Hypotheses**

Hyaluronan-cholesterol nanohydrogels bind to hyaluronidase and halt its enzymatic activity, opening new avenues for the development of HAase-resistant injectable formulations both for biomedical and dermo-cosmetic applications.

## 2. Experimental

### 2.1. Materials

Hyaluronan tetrabutylammonium salt (HA$^-$TBA$^+$, $M_w = 2.2 \times 10^5$) and hyaluronan sodium salt (HA$^-$Na$^+$, $M_w = 7.0 \times 10^5$) were purchased from Contipro (DolníDobrou, Czech Republic). Gellan Gum tetrabutylammonium salt (Ge$^-$TBA$^+$, $M_w = 2.5 \times 10^6$) was kindly provided from Novagenit (Trento, Italy). Jonexa$^{TM}$ Hylastan SGL-80 is a commercial product. Hyaluronidase from bovine testes (451 units/mg), cholesterol (CH), N-acetyl-D-glucosamine (N-Ac-D-Glu), 4-bromobutyric acid, N-methyl 2-pyrrolidone (NMP), N-(3-dimethylaminopropyl)-N′-(ethylcarbodimide hydrochloride) (EDC·HCl), acetonitrile (HPLC grade × 99.8%), phosphate buffered saline tablets (PBS), trypsin-ethylenediaminetetraacetic acid (trypsin-EDTA, solution 10×), L-glutamine solution, sterile-filtered water suitable for cell culture, fetal bovine serum (FBS), gelatin from bovine skin, sodium chloride, acetone, sodium nitrate, 4-(dimethylamino)pyridine (DMAP), dichloromethane (HPLCgrade ×99.8%), boric acid, dextrose, potassium hydroxide, 4-(dimethylamino)benzaldehyde (DMAB), ammonium acetate and cyclohexane (reagent grade ×99.5%) were purchased from Sigma-Aldrich (Milan, Italy). Purified hyaluronidase from Bovine testes was purchased from Worthington Biochemical Corporation (New Jersey, USA). Dulbecco's Modified Eagle Medium (DMEM 1×) was purchased from Gibco brl life technologies Inc., Grand Island (NY, USA). Nitrotetrazolium blue chloride (MTT, CellTiter 96 Non-Radioactive Cell Proliferation Assay) was purchased from Promega (Milan, Italy). Ethyl acetate, sodium sulfate anhydrous, hydrochloric acid 37% and glacial acetic acid were purchased from Carlo Erba (Milan, Italy).

### 2.2. General cell culture

Human Keratinocytes (HaCaT) were spontaneously immortalised from primary keratinocytes[22] and cultured in high glucose DMEM (EuroClone, Milan, Italy) supplemented with 10% (v/v) fetal bovine serum (FBS), 4 mM L-glutamine and 50 µg mL$^{-1}$ gentamicin. Human Dermal Fibroblasts (HDF) were supplied by ATCC (VA, USA) and cultured in DMEM supplemented with 10% FBS, and 2 mM L-glutamine and 1% (v/v) Penicillin-Streptomycin. Human Umbilical Vein Endothelial Cells (HUVEC) were purchased from Promocell (Heidelberg, Germany) and cultured in Promocell Endothelial Cell Growth Medium with SupplementMix (Promocell, Heidelberg, Germany) in a gelatin-coated flask. All cell lines were maintained at 37°C in a humidified atmosphere of 5% $CO_2$, grown to 70-80% semi-confluence (according to each experimental setting) and treated with HA-CH$_{11}$ MHs at several incubation times and concentrations. For each experiment, cells treated with PBS were used as negative control.

2.3. Synthesis of cholesterol-Br-butyric derivative (CH-Br)

CH-Br synthesis was carried out as previously reported[16]. Briefly, 500 mg of cholesterol (CH) (1.3 mmol) were solubilised in 5 mL of $CH_2Cl_2$ and added to 79 mg of DMAP (0.65 mmol); the solution was stirred for 15 min at 25°C. Meanwhile, 648 mg of 4-bromobutyric acid (3.9 mmol) and 744 mg of EDC·HCl (3.9 mmol) were solubilised in 5 mL of $CH_2Cl_2$; the two solutions were then mixed and left at 25°C with magnetic stirring, overnight. The reaction mixture was monitored until completion, by spotting on silica gel TLC (cyclohexane : ethylacetate, 85:15) and then washed once with 50 mM NaOH and HCl and three times with bi-distilled water. The solution was firstly dried with anhydrous $Na_2SO_4$ and then the solvent was removed under vacuum with a rotary evaporator (Heidolph, Heizbad Hei-Vap, Germany). The crude product (powder) was finally purified with a silica column (eluent, cyclohexane : ethylacetate, 98.5:1.5), yielding 60% (w/w) of a white solid.

2.4. Synthesis of hyaluronan-cholesterol derivatives (HA-CH)

HA-CH synthesis was carried out as described in a previous work[16]. 500 mg of HA$^-$TBA$^+$ (323 µmol of repeating unit, $M_w$ = 2.2 x 10$^5$) were added to 10 mL of NMP and the solution was left at 25°C for 5 h under magnetic stirring; then, 65 or 39 mg of CH-Br (corresponding to 121 or 73 µmol), previously solubilised in 2 mL of NMP, were added to HA$^-$Na$^+$ solution and the reactions were allowed to proceed for 48 h at 38°C under magnetic stirring. Then, 2 mL of NaCl (saturated solution) were added drop by drop to the mixtures, which were left under stirring for 30 min to allow the replacement of TBA$^+$ ions with Na$^+$ ions. The reaction products were precipitated in acetone:water (90:10, v/v) (4 times the reaction volume), left for 1 h at 4°C, isolated, dispersed in bi-distilled water and finally dialysed against water (cellulose membrane tubing, molecular weight cut-off: 1.2-1.4 x 10$^4$ Da, Sigma-Aldrich, Darmstadt, Germany) until constant conductivity was reached. Samples were freeze-dried with a "Modulyo 4K" Edwards High Vacuum instrument, equipped with an Edwards pump, yielding 350 mg of white solid (70% mass recovery). $^1$H-NMR analysis showed a degree of functionalisation (DF) of 11 or 5% (mol/mol, i.e., moles of CH per mole of HA repeating unit) (Fig. 1SI); reaction yield was 73 or 56%, respectively. Spectra were recorded on 0.5% wt. polymer solutions in a 1:4 mixture of D$_2$O : *d$_6$*-DMSO using a 400 MHz Bruker spectrometer (Bruker UK Limited, UK). Samples were dissolved by sequential addition of D$_2$O (5 h stirring) and *d$_6$*-DMSO (overnight stirring) to allow adequate solubilisation of the polymers. Spectra were recorded at 50°C with 128 scans in order to obtain a good signal resolution and a high signal-to-noise ratio. Recovery time (aq + d1) of 14 and 64 s were respectively selected to allow full recovery of magnetisation after each 30° pulse. The DF was calculated from spectra at 50°C by means of the integration of the peaks at 4.78-4.09 ppm (H3 + anomeric protons of hyaluronic acid) and 0.59 ppm (CH$_3$ 18 group from cholesterol) using an iterative method for the estimation of the signal integrations. The proton identification in the HA-CH spectrum was realised by means of $^1$H-$^1$H COSY experiment and from data reported in the literature[23].

2.5. Synthesis of gellan-cholesterol derivative (Ge-CH)

A Ge-CH conjugate with a theoretical DF of 20% was prepared as previously described[24, 25]. Firstly, the MW of Ge$^-$TBA$^+$ was reduced through an extrusion process, carried out using an extruder M-110EH-30 Microfluidizer® Processor. Specifically, the polymer aqueous solution (0.5%, w/w) was introduced in the extrusion tube G10Z (87 m) and the process was carried out at 1,200 bar at 50°C. Seven extrusion cycles were performed. The resulted Ge$^-$TBA$^+$ was then functionalised with CH-Br. Briefly, 200 mg of Ge$^-$TBA$^+$ (corresponding to 225 mol) were added to 16 mL of NMP and left under magnetic stirring for 2 h at 25°C. Meanwhile, 24 mg CH-Br (corresponding to 45 µmol) were solubilised in 5 mL of NMP and then added to Ge$^-$TBA$^+$; reaction was allowed to proceed for 40 h at 38°C. The resulting Ge-CH was dialysed against water (molecular weight cut-off: 1.2-1.4 x 10$^4$ Da) until constant conductivity was reached. Samples were freeze-dried, yielding 170 mg of a white solid (85% mass recovery). FT-IR spectrum of Ge-CH$_{20}$ was recorded on a Perkin-Elmer Spectrum One instrument equipped with ATR system.

2.6. Formation of HA-CH$_{11}$ or Ge-CH$_{20}$ NHs

4 mg of HA-CH$_{11}$ or Ge-CH$_{20}$ were added to 2.7 mL of bi-distilled water. The suspensions were left overnight under magnetic stirring at 25°C and then autoclaved at 121°C for 20 min[26], leading to the formation of NHs[12, 22]. The nano-suspensions were finally freeze-dried and stored as dry powder.

2.7. Preparation of hybrid HA mixtures (HHs)

50 mg of HA-CH$_{11}$ were dispersed in 33.3 mL bi-distilled water at 25°C and left overnight under magnetic stirring; samples were then autoclaved at 121°C for 20 min and freeze-dried. Meanwhile, 15 mg HA (HA$^-$Na$^+$, M$_w$ = 7.0 x 10$^5$) were solubilised with 1 mL of bi-distilled water under magnetic stirring (25°C, overnight). Then, HA-CH$_{11}$ NHs were re-suspended (by vortexing for 5 min) with 0.5 mL bi-distilled water at concentration ranging from 0.3 to 3% w/v, and added drop wise over 1 h to a HA solution (1.5% w/v) under magnetic stirring at 25°C (final concentration of HA-CH$_{11}$ NHs ranged from 0.1 to 1% w/v, whilst the final concentration of HA was kept constant at 1% w/v for each mixture). As controls, HHs based on HA-CH$_5$ (which does not form NHs) or Ge-CH$_{20}$ NHs were also prepared by following the same procedure.

Furthermore, HHs were prepared with several weight ratios of HA and HA-CH$_{11}$ NHs, where the total and final HA concentration (native HA + HA-CH$_{11}$) was fixed to 1% w/v. An adequate amount of HA, ranging from 0 to 15 mg, was dissolved in 1 mL of bi-distilled water and respectively mixed with an adequate amount of HA-CH$_{11}$ (ranging from 0 to 15 mg) dispersed in 0.5 mL bi-distilled water (corresponding to a weight ratio ranging from 100:0 to 0:100 (HA:HA-CH$_{11}$)). All samples were added to 22.5 µL 1X PBS (pH 6.2).

2.8. Preparation of physical hydrogels (PHs)

90 or 105 mg of HA-CH$_{11}$ were dispersed in 60 or 70 mL of bi-distilled water, respectively, magnetically stirred overnight at 25°C and then autoclaved at 121°C for 20 min. Suspensions were freeze-dried and then re-suspended in 1.5 mL of bi-distilled water (corresponding to 6 and 7 % w/v, respectively) and added to 67.5 µL of 20% w/v NaCl solution (final concentration of 0.9% w/v). Hydrogels were centrifuged at 6,000 rpm for 10 min at 25°C in order to remove bubbles and then added to 52 µL 1X PBS (pH 6.2).

In Table 1, the composition of HHs and PHs is reported, along with the marketed cross-linked HA.

|  | HA | HA-CH | HA-CH NHs |
|---|---|---|---|
| **HHs** | 1% (w/v), M$_w$ = 7.0x10$^5$ |  | 0.1, 0.5 or 1% (w/v) |
| **PHs** |  | Autoclaved, freeze-dried and re-dispersed HA-CH at 6 or 7% (w/v) |  |
| **Hylastan (commercial cross-linked HA)** | Formed by cross-linked HA particles (~ 20% wt.) and free HA (~ 80% wt.) |  |  |

**Tab.1** Table summarizing the components of the different formulations that have been developed and studied in this work.

2.9. Transmission or Scanning Electron Microscopy (TEM or SEM)

TEM micro-graphs of HA-CH$_{11}$ NHs were obtained using a Philips CM120 instrument. Negative staining with uranyl acetate was employed for better contrast. A droplet of HA-CH$_{11}$ NHs suspension (1 mg mL$^{-1}$, prepared as described in section 2.6) was added onto a 400-mesh copper grids coated with formvar/carbon support film (Agar Scientific1, UK). The excess of sample was blotted off with paper filter. The grid containing the sample was then covered with a small drop of stain solution (1% uranyl acetate dissolved in water, pH 4.5) and the sample allowed to dry at 25°C, before imaging under TEM. The images were captured with a high-resolution digital camera coupled to the microscope.

VP-SEM micro-graphs of PHs were obtained using a variable pressure SEM (Hitachi SU-3500, Italy) with dual energy dispersive X-ray spectroscopy detectors (VP-SEM-dEDS), which were arranged in parallel configuration (Bruker, XFlash®660). PHs were prepared as described in section 2.8 and directly settled onto a carbon planchet stub. Experiments were performed at variable pressure and in conjunction with a cooling stage as well as water vapour in the VP-SEM chamber, in order to stabilise the hydrated sample. Images were recorded with the accelerating voltage of 8 kV, magnification of 1000x and working distance of 5.4 mm.

2.10. HAase treatment of HA-CH$_{11}$ derivative and NHs thereof

In all enzymatic degradation experiments, HAase powder (Sigma Aldrich) was employed without further purification steps and solubilized in water + 1X PBS (pH 6.2).

2.10.1. Gel Permeation Chromatography (GPC) measurements

12 mg of HA or HA-CH$_{11}$ were dissolved in 6 mL of bi-distilled water (2 mg mL$^{-1}$) and left overnight under magnetic stirring at 25°C. Then, 2 mL of sample were added to 0.293 mL of 1 mg mL$^{-1}$ HAase in water (final EU mL$^{-1}$ = 66) and 32 μL 1X PBS (pH 6.2). Enzymatic reactions were left for 2 or 6 h at 25°C without shacking. As controls, HA or HA-CH$_{11}$ without HAase were also prepared and added to 32 μL 1X PBS (pH 6.2). Samples were freeze-dried, solubilised in bi-distilled water (HA, 1.7 mg mL$^{-1}$) or DMSO (HA-CH$_{11}$, 2.2 mg mL$^{-1}$), filtered 0.22 μm and finally injected (injection volume = 200 μL) into GPC apparatus. Molecular weight distribution ($M_n$ and $M_w$), polydispersity (PD), hydrodynamic radius ($R_h$) and the value of the Mark-Houwink-Sakurada equation of HA and HA-CH$_{11}$, before and after de-polymerisation by HAase, were determined by using a Viscotek Triple

Detector Array (Viscotek TDA 305) equipped with Refractive Index (RI), Light Scattering (RALS and LALS) and Viscosity detectors. For the analysis of HA samples, an aqueous mobile phase of 100 mM $NaNO_3$ supplemented with 0.01% w/v $NaN_3$ was employed, at the flow rate of 0.6 mL min$^{-1}$, by using three serial columns (TSK gel 13 µm, 300x7.8 mm, 100-1000 Å) at 35°C. The instrument calibration was performed with poly(ethylene oxide) (PEO, $M_w$ = 2.4 x $10^4$) as a narrow standard and validated with a polydisperse dextran ($M_w$ = 7.3 x $10^4$) as broad standard. In the case of HA-$CH_{11}$ samples, the analyses were performed in DMSO at 60°C, at flow rate of 0.6 mL min$^{-1}$ using two serial columns (Phenogel 5 µm, 300 x 7 mm, 500 and $10^5$ Å, respectively). The instrument calibration was performed with pullulan ($M_w$ = 1.05 x $10^5$) as a narrow standard and validated with a polydisperse dextran ($M_w$ = 7.3 x $10^4$) as broad standard. In order to estimate the refractive index increment (dn/dc) values of the polymers, filtered samples (0.22 µm pore size) at concentration ranging from 1 to 4 mg mL$^{-1}$ were sequentially injected into the GPC apparatus and analysed assuming 100% mass recovery. Signals were analysed using OmniSEC 5.0 software (Malvern). All samples were analysed in triplicate.

2.10.2. Dynamic Light Scattering (DLS) measurements

4 mL of HA-$CH_{11}$ NHs suspensions were prepared as described in section 2.6. 1 mL sample was added to 0.146 mL of 1 mg mL$^{-1}$ HAase aqueous solution (final EU mL$^{-1}$ = 66) and 16 µL 1X PBS (pH 6.2). Enzymatic reactions were left for 2 or 6 h at 25°C, without shaking. As control, a blank of HA-$CH_{11}$ NHs, without HAase, was prepared. Hydrodynamic diameter (Z-average size), size distribution, polydispersity (PDI) and count rate of 1 mg mL$^{-1}$ HA-$CH_{11}$ NHs suspensions, before and after HAase treatment, were measured with DLS at 25 °C by using a Zetasizer Nano ZS instrument (Model ZEN3690, Malvern Instruments) equipped with a solid state HeNe laser ( = 633 nm) at a scattering angle of 173°. Size measurement data were analysed by using the general-purpose algorithm. Each experiment was performed in triplicate.

2.11. Degradation of HHs *vs* Hylastan by HAase

1.5 mL of HHs were prepared as described in section 2.7. Each HH was added to 100 µL HAase water solution with concentration ranging from 2 to 8 mg mL$^{-1}$ (corresponding to 66-264 EU mL$^{-1}$, respectively) for 2 h, without magnetic stirring, at 25°C. Flow curves were recorded at 25°C, using a controlled stress rheometer (Anton Paar MCR102), by using a cone-plate geometry (diameter = 50 mm; cone = 2°). Flow curves were determined in the range of 0.001-1000 s$^{-1}$ at 25 °C. A stepwise increase of the stress was applied with an equilibration time of 30 s. Each gel was allowed to rest for 5 minutes before starting the measurements. All samples were analysed in triplicate.

HAase treatment of HHs was also compared to that of Hylastan over the time. Specifically, 90 mg HA-CH$_{11}$ were dispersed in 60 mL of bi-distilled water under magnetic stirring at 25°C, overnight. Then sample was autoclaved at 121°C for 20 min and freeze-dried. 15 mg HA (HA$^-$Na$^+$, M$_w$ = 7.0 x 10$^5$) were solubilised in 1 mL of bi-distilled water under magnetic stirring at 25°C, overnight. Then, freeze-dried NHs were re-suspended in 0.5 mL bi-distilled water, by vortexing for 5 min at concentration of 0.3 or 3% w/v, and added drop wise over 1 h to 1 mL of HA solution (1.5% w/v), under magnetic stirring at 25°C. Final concentration of HA-CH$_{11}$ corresponded to 1 or 0.1% w/v, whilst the final concentration of HA was 1% w/v in all samples. Each mixture was then added to 29.3 μL HAase water solution (15 mg mL$^{-1}$) at day 0, 2, 5 without magnetic stirring, at 25°C in the presence of 20 μL of a 2 mM Na$_3$N aqueous solution. All samples received 22.5 μL of 1X PBS (pH 6.2). Flow curves were recorded at 25°C, at days 0 (without HAase), 1, 2, 3 and 7; the final volume of each sample was 1.7 mL. The commercial gel, Hylastan, received the same HAase treatment.

2.12. Degradation of PHs *vs* Hylastan by HAase

1.5 mL of 6 or 7% w/v PHs were prepared as described in section 2.8 and added to 29.3 μL of HAase water solution (15 mg mL$^{-1}$, corresponding to a final concentration of EU mL$^{-1}$ = 132). PHs were left at 25°C without stirring for 24 h, then 29.3 μL of HAase water solution (15 mg mL$^{-1}$) were added again and samples were left for other 24 h at 25°C. As comparison, Hylastan was added to 22.5 μL 10X PBS (pH 6.2) and undergone to the same HAase treatment. Negative controls consisted of PHs without HAase were prepared following the same procedure. On the third day, mechanical spectra and flow curves were recorded at 25°C, using the rheometer. Prior to recording rheological data, stress sweep measurements were performed at 25°C and 1 Hz to assess the linear viscoelastic region of gels. Mechanical spectra were then recorded at 25°C in the range of 0.01-100 Hz, applying a constant deformation ( = 0.01). Flow curves were determined in the range of 0.001-1000 s$^{-1}$ at 25 °C. A stepwise increase of the stress was applied with an equilibration time of 30 s. Each gel was allowed to rest for 5 minutes before starting the measurements. All samples were analysed in triplicate.

2.13. Study of the inhibition kinetics of HAase enzymatic activity

2.13.1 Reissig assay

4.94 g of boric acid and 1.98 g of potassium hydroxide were added to 100 mL of bi-distilled water and the solution was left under magnetic stirring at 25°C. 385 mg of ammonium acetate were added to 1 L of bi-distilled water (5 mM) and pH was adjusted to 5 with either acetic acid or ammonia solution. 0.5 g of p-dimethylaminobenzaldehyde (DMAB) were added to 0.625 mL of HCl 12 N and 4.375 mL of glacial acetic acid. DMAB solution (100 mg mL$^{-1}$) was diluted 10-fold with glacial acetic acid just before use (at least 15 min prior). N-Ac-D-Glu was added to 5 mM ammonium acetate solution (pH 5) at a concentration ranging from 67.5 to 550 µg mL$^{-1}$ (1.5 mL) and used as a standard. 1 mg mL$^{-1}$ of HA ($M_w$ = 2.2 x 10$^5$) solution was prepared in ammonium acetate (5 mM, pH 5, 1.5 mL) solution and left under magnetic stirring for 2 h at 25°C. Meanwhile, 5 mg of HAase were added to 0.625 mL of ammonium acetate (8 mg mL$^{-1}$) solution and the reaction was started with the addition of 100 µL of HAase solution to 1.5 mL HA or N-Ac-D-Glu solutions at 37°C. After 7 min, 200 µL of reaction were added to 50 µL of boric acid solution. Samples were left at 100°C for 5 min and then cooled at 4°C for 5 min. Then, 1.5 mL of diluted DMAB solution were added and the reactions were left at 37°C for 15 min. Samples were immediately scanned from 700 to 400 nm with a Perkin-Elmer double beam õLambda 3Aö UV-Vis spectrometer. Analyses were performed at 25°C, using 1 mm quartz cuvettes (Hellma Analytics, Milan, Italy), at 585 nm. The calibration curve was built with N-Ac-D-Glu at concentration ranging from 67.5 to 550 µg mL$^{-1}$ ($R^2$ = 0.992, n = 5). The enzymatic activity of HAase was calculated by using a range of solutions with increasing HAase final concentrations (from 0.33 to 1.7 mg mL$^{-1}$) added to 1.5 mL of HA solution. 1 EU was defined as the amount of HAase able to produce an increase in absorbance of 0.001 at 585 nm and 37°C in 1 min. Each experiment was performed in triplicate.

2.13.2 Michaelis-Menten kinetics and inhibition activity

A number of 1.4 mL aliquots of HA (ranging from 0.033 to 1.07 mg mL$^{-1}$) dissolved in 5 mM ammonium acetate (pH 5) were separately added to either 100 µL of HA-CH$_{11}$ NHs aqueous suspensions at concentration ranging from 1.6 to 6.4 mg mL$^{-1}$ (corresponding to final NHs concentrations ranging from 0.1 to 0.4 mg mL$^{-1}$) or 100 µL of bi-distilled water (control). Samples were stirred for 30 min at 25°C. Then 100 µL of HAase solution in 5 mM ammonium acetate (pH 5) were added to each mixture (12 mg mL$^{-1}$, corresponding to 248 EU mL$^{-1}$). Enzymatic reactions were allowed to proceed for 7 min at 37°C without stirring. Then 200 µL of each mixture were removed and the Reissig assay was performed for each sample (as described above). UV-Vis readings were immediately recoded at 585 nm, in order to assess the HAase enzymatic activity. As controls, NHs suspensions with or without HAase, as well as HA without HAase were assayed. Michaelis-Menten and Lineweaver-Burk curves were fitted with Origin Pro 2016 and GraphPad Prism 7 Software, from which the Michaelis-Menten constant ($K_m$), $K_{m(app)}$, the maximum velocity ($V_{max}$), $V_{max(app)}$ and the inhibition constant ($K_i$) values were calculated.

2.14. Binding of HA-CH$_{11}$ NHs to HAase

2 mg of HA-CH$_{11}$ NHs or Ge-CH$_{20}$ NHs or HA were added to 2 mL of 1X PBS (1 mg mL$^{-1}$) and left under magnetic stirring at 25°C, overnight. Samples were degassed and titrations were performed at 25 °C, during which HAase solution (71 μM) was placed into the syringe injector, whereas 200 μL of each polymer (2-2.5 μM) were loaded in the cell, independently. Isothermal titration calorimetry (ITC) measurements were performed by using MicroCal ITC200 microcalorimeter (MicroCal Inc., Northampton, MA, USA; Malvern). The titrations involved 19 injections of 2 μL at 150 s intervals. The stirring speed of the syringe was set at 400 rpm. Reference titrations of HAase were used to correct the heats of dilutions. The thermodynamic data were processed with Origin 7.0 software that was provided by MicroCal. The enthalpy change (ΔH) values were measured for each titration, by fitting the binding isotherms with a one-site binding model, yielded the values of the association constant ($K_a$). Measurements also gave information about the entropy change (ΔS). The binding free energy (ΔG) and the dissociation constant ($K_d$) were calculated from the experimentally determined values of ΔH and Ka, by using eq. 1 and 2:

$$\Delta G = -RT \ln K_a = \Delta H - T\Delta S \quad (1)$$

$$K_d = \frac{1}{K_a} \quad (2)$$

where R is the gas constant (1.987 cal·mol$^{-1}$·K$^{-1}$) and T is the working temperature (298 K). Results are the average of triplicate measurements. Baseline adjustment was performed manually to correct any discrepancies in the baseline outlined by the software. The $K_a$ parameter was fixed when necessary (typically at the beginning of the curve), in order to obtain a better fitting of the data.

2.15. Cell viability assay

HaCaT (5,000 cells/well in complete DMEM), HUVEC (10,000 cells/well in Promocell Endothelial Cell Growth Medium, gelatin- coated plate) and HDF (2,500 cells/well in DMEM) were seeded in a 96-well plate (Falcon Polystyrene Microplates, Thermo Fisher Scientific, Monza, Italy) in 100 μL of complete medium and incubated for 24 h under the cell culture conditions explained in section 2.2. Subsequently, 25 μL of each sample (HA-CH$_{11}$ NHs in PBS pH 7.4) at specific final concentrations (ranging from 18 to 500 μg mL$^{-1}$) were added to each well, and then the cells were incubated for 24 and 48 h. As negative control, cells received 25 μL of PBS. Then, medium was removed, cells were gently washed with PBS and 100 μL complete DMEM were added to each well. 20 μL MTT solution (Promega, Italy, Milan) were added to each well and cells were incubated for 2 h. Supernatants were gently removed and formazan crystals were solubilized with 100 μL DMSO. Absorbance was checked at 570 nm with a reference at 690 nm, using an Appliskan microplate reader (Thermo Scientific, Vantaa, Finland). Each experiment was performed on 16 wells (n = 3). Results were processed using SkanIt 2.3 Software.

2.16. Statistical analysis

Cell viable values are averages of three biological replicates (each value was calculated from readings of 16 independent wells) and are expressed as the mean value ± standard deviation. Cell viability was normalised to the negative control (untreated cells that received PBS). Statistical significance was determined with One-way ANOVA analysis in Prism (GraphPad 5.0 Software, Inc., La Jolla, CA, USA). Differences between groups were determined by a Turkey's multiple comparison test. Asterisks denote statistically significant differences (*$p < 0.05$; **$p < 0.01$; ***$p < 0.005$).

## 3. Results and discussion

A functional bromoalkyl cholesterol (CH) derivative was conjugated to carboxyl groups of HA through an ester bond[16], producing HA-CH$_{11}$ with a DF of 11% (mol of CH per mol of HA repeating unit) (Fig. S1). The resulting amphiphilic polymer was treated with a sterile autoclaving cycle (121°C, 20 min)[12, 26] to produce NHs suspensions (at the concentration of 0.1% w/v) with a mean hydrodynamic diameter of 250 ± 30 nm, low PDI (0.18 ± 0.05) and highly negative ζ-potential (-40 ± 3 mV). Such NHs show high stability in aqueous environment (for months at 4°C) and can be freeze-dried and stored as a powder, preserving their starting hydrodynamic diameter and PDI after re-hydration. The physico-chemical properties of NHs (e.g. size distribution, morphology, ζ-potential and critical micelle concentration) as well as the effect of the temperature upon the NHs size and PDI, have been described in detail in previous works[12,16]. The stability against the enzymatic degradation of HA-CH$_{11}$ either in suspension (Fig. 1B) or as NHs thereof (Fig. 1C) was initially assessed by monitoring GPC and DLS changes of samples, treated with HAase (final EU mL$^{-1}$ = 66) for 2 or 6 h. Surprisingly, both HA-CH$_{11}$ polymer and its NHs showed negligible changes by both techniques (i.e. little-to-none enzymatic degradation) whilst, as expected, for pristine HA a drastic molecular weight ($M_n$ and $M_w$) decrease (ca. 80% reduction after 6 h) was detected (Fig. 1A). It is worth noting, the HAase employed herein is from bovine testes, which should have a primary sequence homologous to that of all human HAases and to that of bee venom HAase[27]; their 3D structures appear also to be similar to one another. Therefore, it is assumed human HAases degrade HA by the same mechanism as other vertebrate HAases[28].

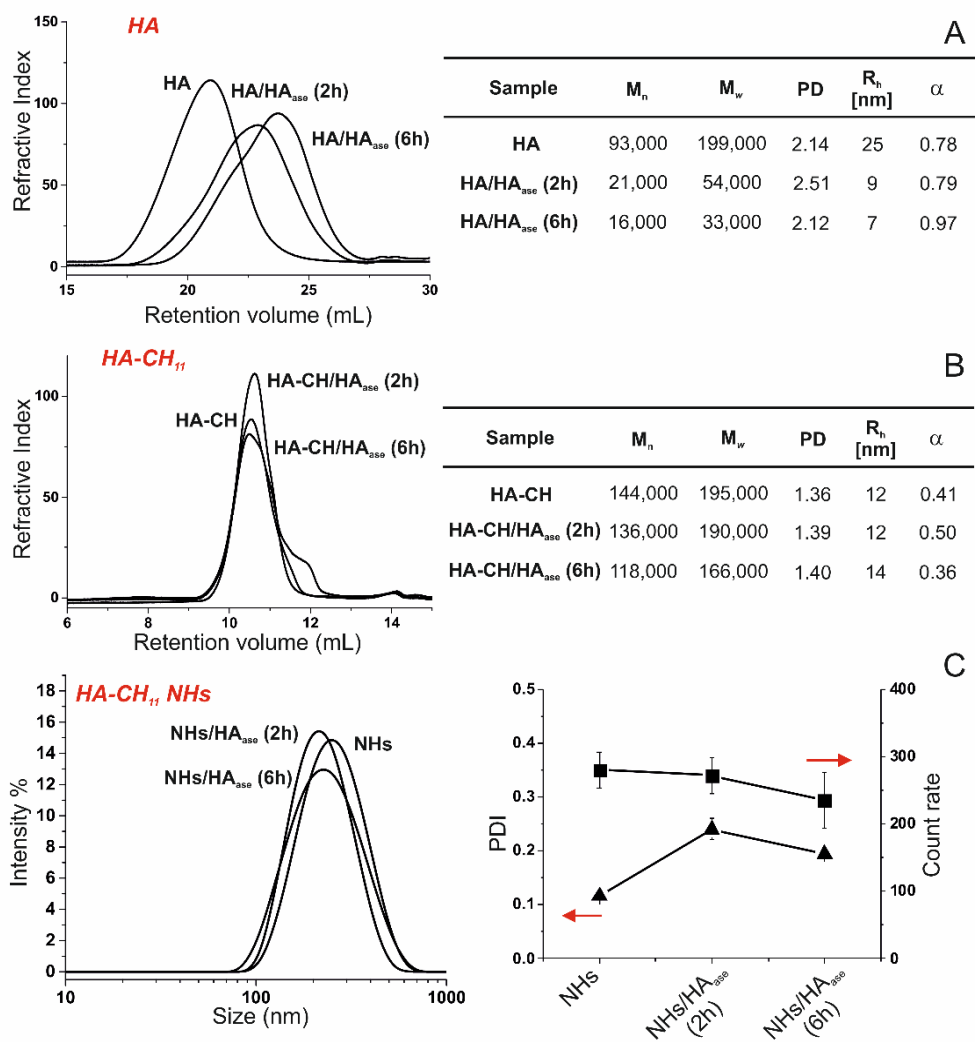

**Fig. 1.** GPC traces, $M_n$, $M_w$, polydispersity (PD), hydrodynamic radius ($R_h$) and α value of the Mark-Houwink-Sakurada equation of (A) HA (1.7 mg mL$^{-1}$) and (B) HA-CH$_{11}$ (2.2 mg mL$^{-1}$) before and after HAase treatment (final EU mL$^{-1}$ = 66) for 2 or 6 h. GPC analyses were carried out in aqueous (HA samples) or DMSO solvents (HA-CH$_{11}$ samples). For the GPC experiments, the difference in the retention times is due to the use of different GPC columns for the two types of samples. (C) DLS size distribution, polydispersity index (PDI) and derived count rate of HA-CH$_{11}$ NHs (1 mg mL$^{-1}$) before and after HAase treatment (final EU mL$^{-1}$ = 66) for 2 or 6 h. Data are expressed as the mean value ± standard deviation (n=3).

Hybrid HA solutions (HHs) prepared by mixing several concentrations of HA-CH$_{11}$ NHs (ranging from 0 to 1% w/v) and 1% w/v HA solution (HA$^-$Na$^+$, Mw = 7.0 x 10$^5$) were subsequently developed and investigated. Fig. 2A clearly shows the presence of NHs slightly affects the viscosity values of HA at all the tested concentrations, leading to the reduction of the zero-shear viscosity, without compromising the pseudoplastic behaviour. ~~HA microparticles sh are able to modify the rheological properties of HA hydrogels[29]~~.

Moreover, the viscosity values of HHs decreased as the concentration of HA was reduced, even though the total HA content (HA + HA-CH$_{11}$ NHs) remained constant (Fig. 2B). Interestingly, such behaviour would allow NHs to be incorporated into HA solutions, without significantly increasing the overall viscosity of the mixtures. As such: I) injectable formulations with similar viscoelastic properties as HA solutions can be developed; II) the viscosity of HHs can be adjusted by turning the MW or the concentration of HA, in order to obtain formulations suitable for viscosupplementation or dermo-cosmetic applications.

We investigated the Stress-relaxation beahvour ph HHs compared to native HA solutioní . (Fig. S2)

The rheological behaviour of HHs was exploited as a probe to study the stability of HHs against HAase degradation, as the depolymerisation of the HA chains would lead to less viscous samples over time. Not surprisingly, when HHs were treated with HAase (final EU mL$^{-1}$ = 66) for 2 h, a negligible decrease of the viscosity values was seen (Fig. 2C), evidencing no HA degradation due to halting of the enzyme activity by HA-CH$_{11}$ NHs.

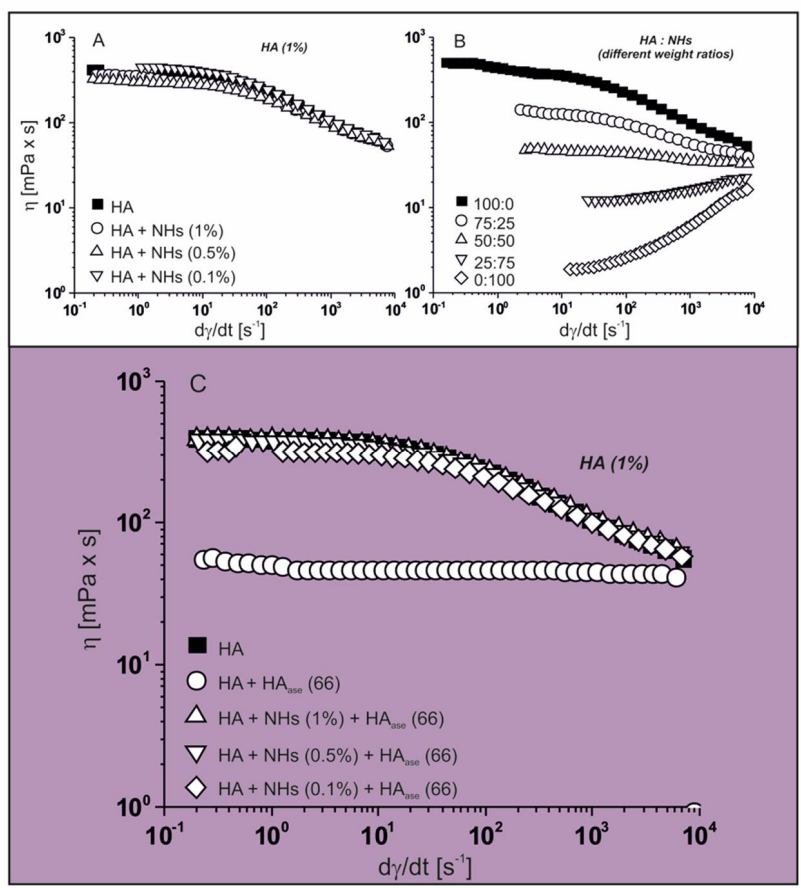

**Fig. 2.** (A) Flow curves of HHs prepared with a fixed HA concentration (1% w/v) and several HA-CH$_{11}$ NHs concentrations (ranging from 0 to 1% w/v). (B) Flow curves of HHs prepared with several ratios of HA and HA-CH$_{11}$ NHs, where the total and final HA content (HA + HA-CH$_{11}$ NHs) was fixed to 1% w/v. (C) Flow curves of HHs prepared with a fixed HA concentration (1% w/v) and several HA-CH$_{11}$ NHs concentrations (ranging from 0 to 1% w/v) after HAase treatment (final EU mL$^{-1}$ = 66) for 2 h at 37°C. As control, flow curves of HA solutions, before and after HAase treatment, are also reported (n=3).

With the aim of assessing HA-CH$_{11}$ NHs as component for injectable formulations, HHs were further studied at two NHs concentrations (0.1 and 1% w/v) and several HAase concentrations (ranging from 66 to 264 EU mL$^{-1}$). A more hydrophilic HA-CH conjugate with a DF of 5% (i.e., HA-CH$_5$), which unlike HA-CH$_{11}$ does not form NHs in aqueous environments, was used as control. Fig. 3A and B clearly show HAase affects each of these systems in a similar manner: I) the viscosity of native HA decreased in a proportional fashion to the HAase concentration; II) HHs containing either HA-CH$_{11}$ NHs or soluble HA-CH$_5$ at 0.1 % (w/v) were only slightly degraded at the highest HAase concentration; III) HHs containing either HA-CH$_{11}$ NHs or soluble HA-CH$_5$ at 1 % (w/v) were not significantly degraded at all the tested HAase concentrations. Fig. 3C confirms the HAase activity mainly depends on the molar concentration of CH moieties linked to the HA chains. Further, at the highest HA-CH$_5$ concentration (1% w/v), the overall viscosity of HHs increases significantly compared to HA-CH$_{11}$ NHs (Fig. 3B, white bars), evidencing only the soluble HA-CH$_5$, but not HA-CH$_{11}$ NHs, adds on significantly to the viscosity of HA. Aggiungere nuove misure e cambiare figura (un numero è sbagliato).

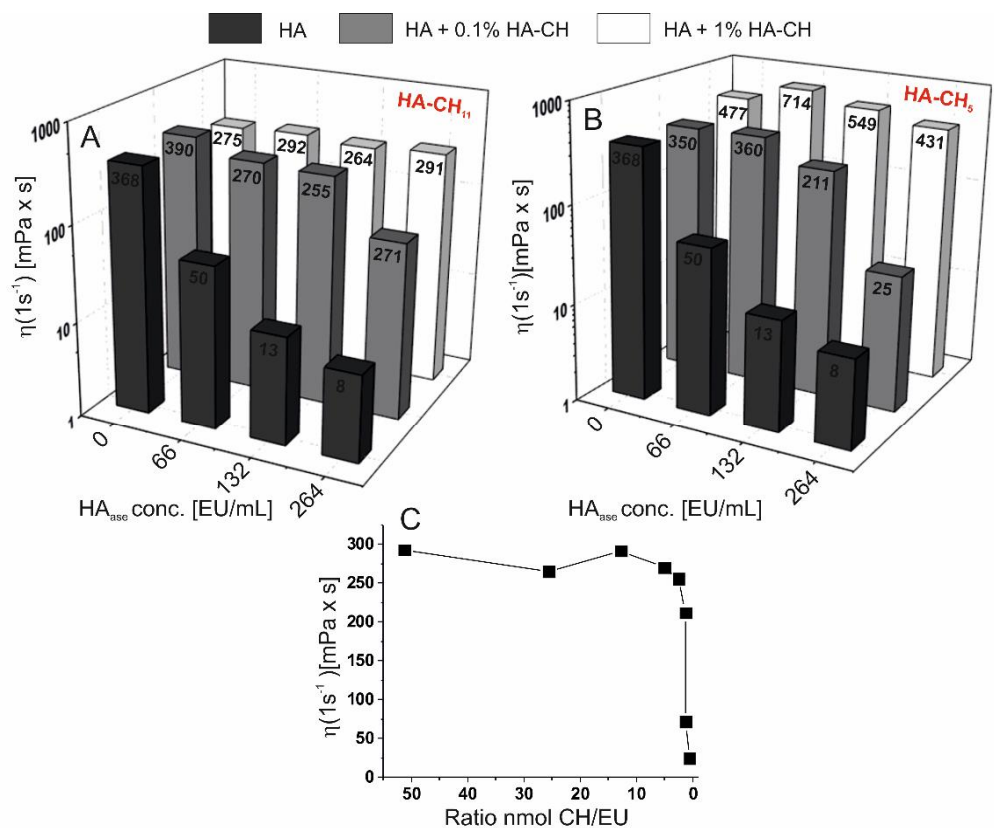

**Fig. 3.** Viscosity values ( ) at 1 s$^{-1}$ of (A) HA-CH$_{11}$ NHs-based HHs and (B) HA-CH$_5$-based HHs with fixed HA concentration (1% w/v) and several HA-CH concentrations (corresponding to 0, 0.1 and 1% w/v) before and after HAase treatment. HAase was added at several final EU mL$^{-1}$ (66, 132 and 264). (C) Graphical representation of HHs viscosity *vs* the ratio between CH (nmols) and HAase (EU), obtained from both HA-CH$_{11}$ and HA-CH$_5$ derivatives (n=3).

An amphiphilic polymer conjugate prepared from another anionic polysaccharide, gellan-cholesterol (Ge-CH$_{20}$), was also analysed (Fig. S3). Ge-CH$_{20}$ forms NHs after autoclaving likewise HA-CH$_{11}$[23]. 1% w/v Ge-CH$_{20}$ NHs were added to free HA to form HHs that were subsequently treated with several HAase concentrations (Fig. S4). Results showed Ge-CH$_{20}$ NHs were highly ineffective in halting

HAase activity as HA degradation occurred readily (reduced suspension viscosity), thus evidencing HAase inhibition is specifically due to the chemical conjugation of CH to HA.

Importantly, HHs were treated with HAase for up to one week (Fig. 4B and C) and their stability against enzymatic degradation compared to that of the commercial supplementation formulation, Hylastan, which is composed of cross-linked HA particles (ca. 20% wt.) and free HA (ca. 80% wt.). Results showed Hylastan (Fig. 4A) was quantitatively degraded after 3 days, whereas 0.1 % HHs were completely degraded only after 7 days. Interestingly, 1 % HHs were scarcely de-polymerised over the whole period of time, evidencing the higher stability of our formulation against *in vitro* HAase degradation, compared to Hylastan. Such result suggests a different interaction mechanism for the cross-linked HA particles in Hylastan and HA-CH$_{11}$ NHs in the HHs; in fact, the former merely slows down the enzymatic degradation of HA, whilst the latter halt the HAase activity, protecting HA from de-polymerisation and this effect is dependent on the NHs concentration.

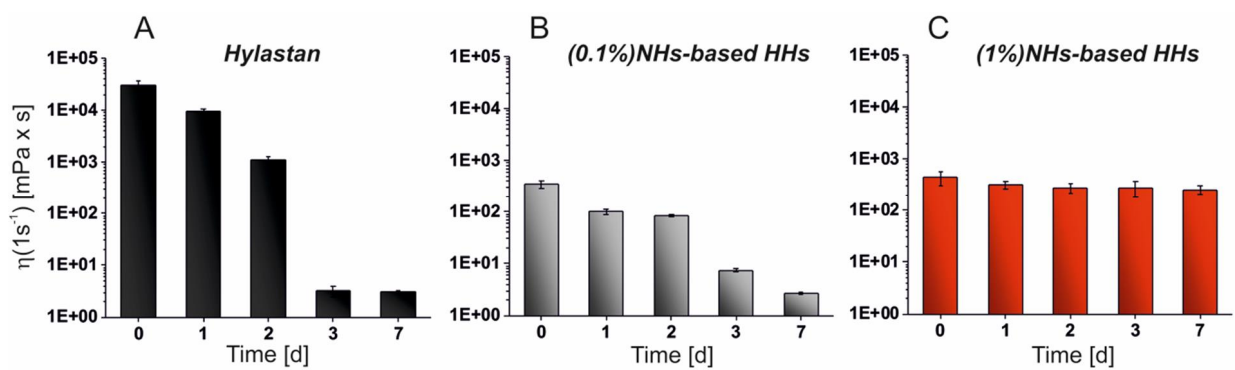

**Fig. 4.** HAase-degradation kinetics of HHs *vs* Hylastan. Viscosity values ( ) at 1 s$^{-1}$ of (A) Hylastan, (B) (0.1%)NHs-based HHs and (C) (1%)NHs-based HHs treated with HAase over one week. Each mixture was added to 29.3 µL HAase at day 0, 2 and 5 at 25°C in the presence of 20 µL of 2 mM NaN$_3$. All samples received 22.5 µL 1X PBS (pH = 6.2) and were analysed at time 0 (without HAase) and after 1, 2, 3 and 7 days. Data are expressed as the mean value ± standard deviation (n=3).

The kinetics of HAase activity in the presence of HA-CH$_{11}$ NHs was studied using the Reissig colorimetric method[30], which is employed here to quantify the reaction products between the Ehrlich's reagent (DMAB) and the N-acetyl-D-glucosamine (N-Ac-D-Glu) reducing end of each HA chain produced after enzymatic scission. Several NHs concentrations (from 0 to 0.4 mg mL$^{-1}$) were employed and the resulting kinetic curves were fitted with both Michaelis-Menten (Fig. 5A) and Lineweaver-Burk (Fig. 5C) models. Results clearly showed NHs halt HAase activity *via* a mixed inhibition mechanism, suggesting NHs can interact HAase both in free state and complexed with HA (HAase/HA), although with different affinities. Indeed, both Michaelis-Menten constant ($K_m$) and maximum velocity ($V_{max}$) significantly decreased (from 0.74 and 0.036 to 0.55 and 0.013, respectively), showing a NHs concentration-dependent effect; the inhibition constant ($K_i$) was found to be 0.342 mM (Fig. 5B).

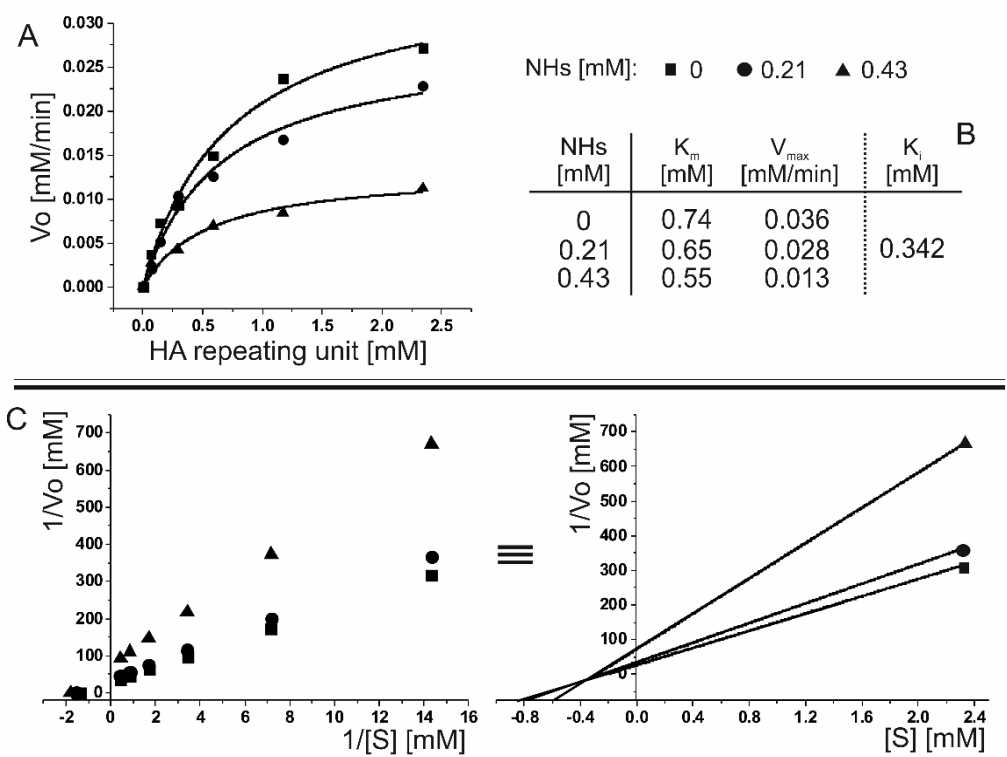

**Fig. 5.** (A) Michaelis-Menten and (C) Lineweaver-Burk fittings. Several HA concentrations (ranging from 0.033 to 1.07 mg mL$^{-1}$) were added to HAase (final EU mL$^{-1}$ = 248) in the presence of NHs (at concentration of 0, 0.1 and 0.4 mg mL$^{-1}$). Subsequently, reactions were tested with Reissig assay and absorbance was recorded at 585 nm. (B) $K_m$, $V_{max}$ and $K_i$ values were calculated with OriginPro 2016 and GraphPad Prism Software. (n=3).

The thermodynamic profiles of the binding between HAase and HA-CH$_{11}$ NHs or free HA evidenced both favourable enthalpic and entropic components. Interestingly, the binding affinity of HA-CH$_{11}$ NHs to HAase was 7.0-folds higher than that of its natural substrate HA, involving a strong favourable enthalpic variation. In contrast, Ge-CH$_{20}$ NHs did not show any affinity for HAase (Fig. 6 and Fig.

S5). Overall, these results evidence an interaction between HAase and its inhibitor, which was different and stronger than that of HA. As HAase active site is lined by a number of cationic and hydrophobic amino acids, it is reasonable to assume the amphiphilic and polyanionic HA-CH$_{11}$ NHs would interact with HAase in a synergistic fashion, where both the presence of HA and CH are fundamental for a strong enzyme inhibition. It is important to point out that although HA-CH$_{11}$ NHs show a preferential disposition of HA in the surface and CH in the interior, some CH chains could be still available to interact with HAase in a similar fashion to the more hydrophilic HA-CH$_5$ conjugate. Moreover, as Pavan and collaborators (2016) showed a certain inhibitory activity of alkyl-HA polymers against HAase[31], this effect may be more generally ascribed to the HA hydrophobisation.

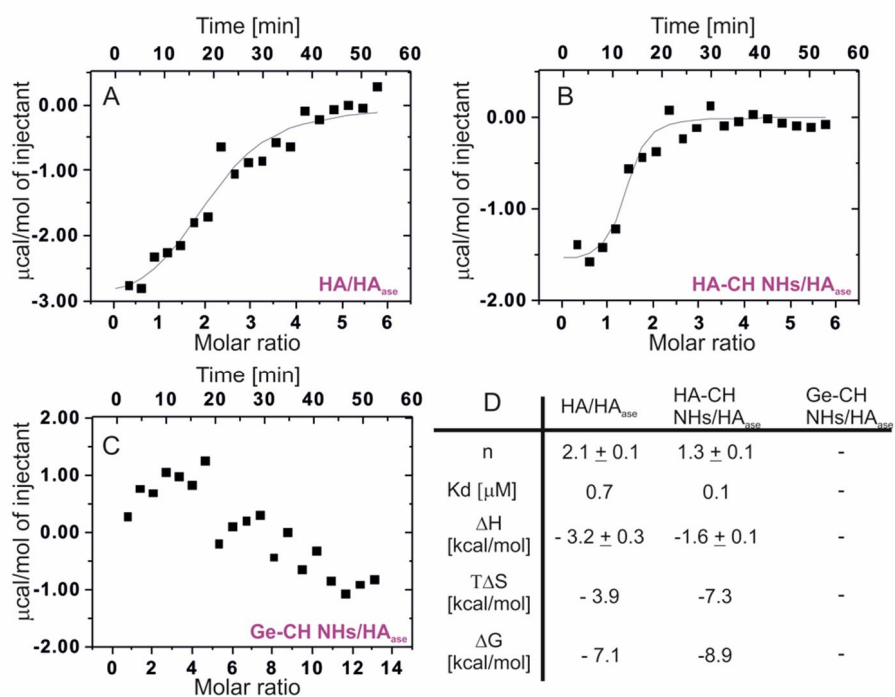

**Fig. 6.** Isothermal titration calorimetry (ITC) graphs of (A) HA, (B) HA-CH$_{11}$ NHs or (C) Ge-CH$_{20}$ NHs and HAase. 200 µL of 1 mg mL$^{-1}$ of each polymer were added to the cell. Every 150 s, 19 individual injections of 2 L of purified HAase (1 mg mL$^{-1}$) were added to the polymer samples. (D) Stoichiometry (n), dissociation constant (K$_d$), enthalpy change ( H), entropy change ( S) and Gibbs free energy ( G) of HA, HA-CH$_{11}$ NHs and Ge-CH$_{20}$ NHs titrations. Each sample was analysed in triplicate. Results were processed with Origin 7.0 Software.

At concentrations largely above 0.1% (w/v), HA-CH$_{11}$ is able to form physically cross-linked networks. Specifically, at concentrations higher than 6% (w/v), hydrated HA-CH$_{11}$ formed physical hydrogels (PHs) in physiological medium (0.9% NaCl, Fig. 8A), which showed a microporous interconnected structure (Fig. 7A), as opposed to NHs formed at 0.1% (w/v) (Fig. 7B)[12, 16]. It should be pointed out, TEM micro-graph shows NHs with smaller size compared to DLS data (Fig. 1C), as the image is recorded in dry state.

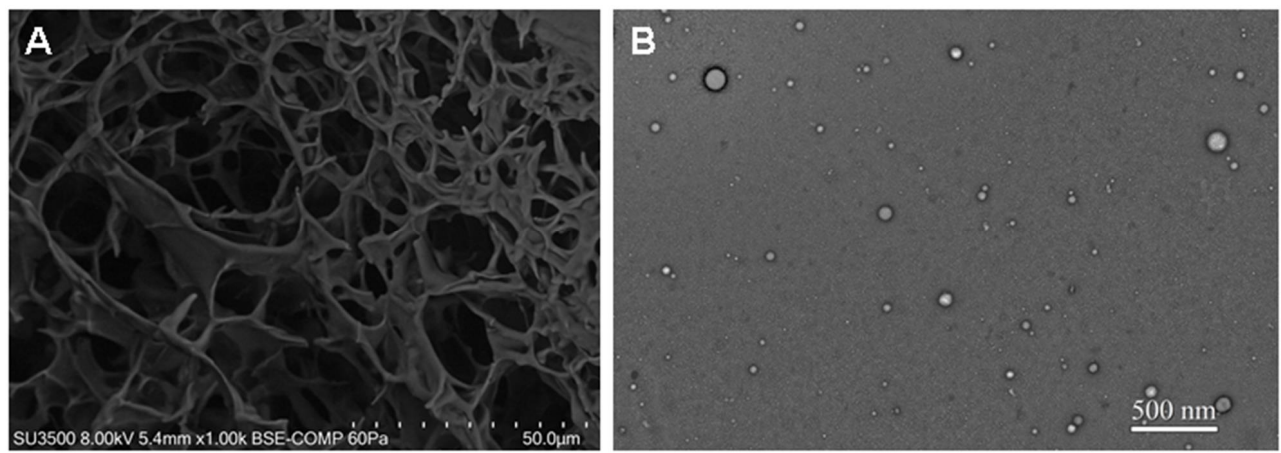

**Fig. 7.** A. VP-SEM micro-graph of porous HA-CH$_{11}$ PH. Autoclaved and freeze-dried HA-CH$_{11}$ was re-hydrated at 6% w/v in physiological medium. B. TEM micro-graph of HA-CH$_{11}$ NHs. Autoclaved and freeze-dried HA-CH$_{11}$ was re-hydrated at 0.1% w/v in aqueous medium.

6 and 7% (w/v) PHs were characterised in terms of mechanical ($G\prime$ and $G\prime\prime$ vs frequency) and shear (viscosity vs shear rate) properties (Fig. 8B and 8C). As comparison, the rheological properties of Hylastan were also studied. Mechanical spectra (Fig. 8B) showed the PHs viscoelastic properties can be controlled by tuning the HA-CH$_{11}$ concentration, i.e., the lower the concentration of HA-CH$_{11}$ the lower the hydrogel elasticity ($G\prime$). Significantly, Hylastan showed a weak gel-like behaviour at the frequency range explored, whereas both PHs samples showed a viscoelastic behaviour above the cross-over point, at 0.02 Hz (Fig. 3B). Furthermore, at the running frequency of 2.5 Hz, the $G\prime$ values of Hylastan, 6% and 7% (w/v) PHs were 58, 33 and 656 Pa, whilst $G\prime\prime$ was respectively 24, 7 and 107 Pa, suggesting the PHs concentration can be adjusted to tailor $G\prime$ and $G\prime\prime$ values. Moreover, all the systems showed comparable viscosities at the shear-rate range usually used for injections (100-300 s$^{-1}$). Flow curves (Fig. 8C) showed Hylastan and PHs have similar zero-shear viscosities as well as similar high shear rate viscosities, even though the slopes of the curves were not superimposable. The difference in viscoelastic behaviour as well as the different mechanical profiles, might relate to a different molecular arrangements of the HA chains/particles within the two systems.

More importantly, 6% (w/v) PH and Hylastan were treated with HAase for 3 days (Fig. 8D): the first system did not show any viscosity change, evidencing no de-polymerisation, whilst the latter was entirely degraded, showing viscosity values close to the water. Such result suggests both PHs and HHs are more resistant to the HAase degradation *in vitro* than the commercial product, Hylastan.

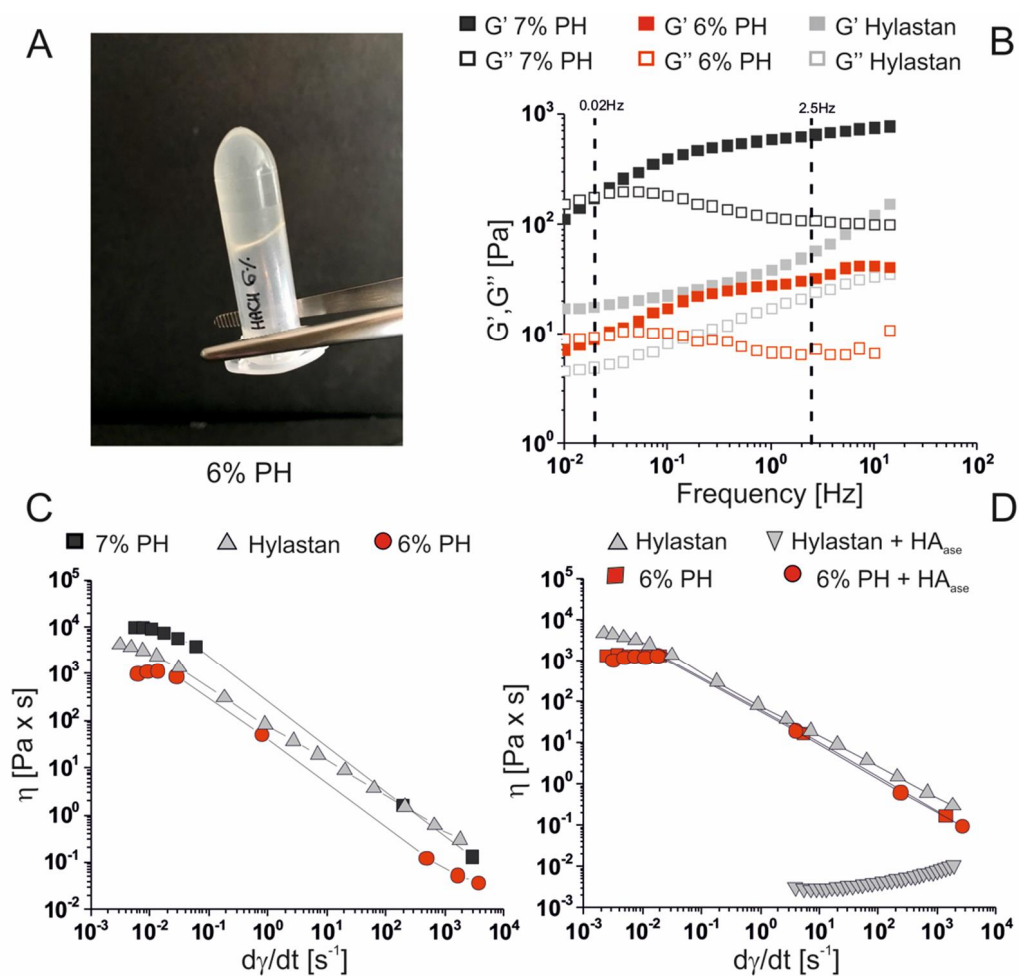

**Fig. 8.** (A) Picture of a 6% (w/v) PH sample. (B) Mechanical spectra (G′ and G″) and (C) flow curves of PHs (6 and 7% w/v) and Hylastan. (D) Flow curves of 6% (w/v) PH and Hylastan before and after treatment with HAase; 29.3 µL of HAase (final EU mL$^{-1}$ = 132) were added at day 0 and 1, at 25°C. Flow curves were recorded at time 0 (without HAase) and after 3 days.

Finally, the toxicity of HA-CH$_{11}$ NHs was assessed by using MTT assay on two cell lines from human skin, HDF and HaCaT, and one cell line from human endothelium, HUVEC, up to 48 h (Fig. 9). MTT assay showed HA-CH$_{11}$ NHs were not toxic to any of these cell lines, at concentrations ranging from 0 to 500 µg mL$^{-1}$, as cell metabolism was not significantly affected over the whole period of time, thus suggesting the employment of HA-CH$_{11}$ systems may represent a novel and versatile approach for formulating not toxic and HAase-resistant injectable materials.

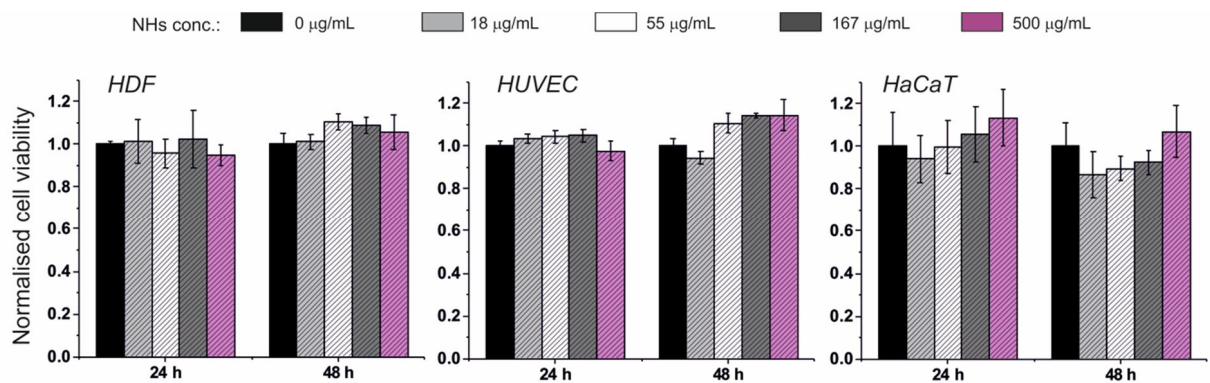

**Fig. 9.** Viability of HDF, HUVEC and HaCaT cell lines upon incubation with HA-CH$_{11}$ NHs. MTT assay was performed on cells incubated for 24 and 48 h with HA-CH$_{11}$ NHs at several concentrations (ranging from 0 to 500 µg mL$^{-1}$). Results were obtained from three independent experiments (each derived from 16 wells). All data are expressed as the mean value ± standard deviation. MTT results were normalised to the control (cells treated with PBS). Statistical significance was determined with One-way ANOVA analysis, GraphPad Prism Software. Differences between groups were determined by a Turkey's multiple comparison test. No significant differences were detected.

4. **Conclusion**

The amphiphilic HA-CH$_{11}$ derivative and assembled NHs thereof clearly showed the ability to turn off the HAase activity, and, hence, halt HA degradation. The extent of such inhibition is related to the amount of CH moles that are linked to the HA chains. As parent Ge-CH$_{20}$ NHs did not inhibit HAase, this inhibitory effect was ascribed to the chemical conjugation of the hydrophobic moiety CH to HA. HAase was inhibited with a mixed mechanism, in which HA-CH$_{11}$ NHs interacted with both free HAase and its complex (HAase/HA), with different affinities, as evidenced by Michaelis-Menten and Lineweaver-Burk models. HAase showed a K$_d$ with its inhibitor that was 7-fold higher than that with its natural substrate, HA; on the contrary, Ge-CH$_{20}$ NHs did not show any binding to HAase. Further studies are necessary to elucidate the specific interaction site between HA-CH$_{11}$ NHs and HAase. As such, HA-CH$_{11}$ NHs were engineered into HHs or PHs systems, which evidenced a number of advantages compared to the currently marketed cross-linked HA (i.e, Hylastan), such as: I) to be more stable to HAase degradation *in vitro*; II) to avoid the use of chemical cross-linkers; III) to increase the injected HA concentration without affecting the overall HA viscosity. Moreover, HA-CH$_{11}$ NHs might be able to entrap certain drugs, i.e. betamethasone, working as drug delivery systems, thus ensuring single injection treatments. Further, taking into account HAase is involved in a number of biological diseases (e.g. tumour development[32] and infections[33]), HAase inhibitors[34] based on nano-systems, could be potentially exploited in different fields in medicine. We believe this novel approach allows improving the performance and biocompatibility of the current injectable HA-based materials for both medical and dermo-cosmetic applications.

**Supplementary material**

Supplementary material available: $^1$H-NMR spectra of HA-CH derivative, Stress-relaxation modulus curves and DSC results of HHs and HA, FT-IR spectra of Ge and Ge-CH$_{20}$, flow curves of HA added by Ge-CH$_{20}$ NHs before and after treatment with several HAase concentrations, ITC graphs of HA, HA-CH$_{11}$ NHs and Ge-CH$_{20}$ NHs added by HAase.

**Author Contributions**

The manuscript was written through contributions of all authors. All authors have given approval to the final version of the manuscript.

**Funding Sources**

Sapienza University of Rome ("Finanziamenti di Ateneo per la Ricerca Scientifica – RP116154C2EF9AC8" and "Progetto di Ricerca RM11715C1743EE89").

**Acknowledgements**

The authors acknowledge financial support from Sapienza University of Rome ("Finanziamenti di Ateneo per la Ricerca Scientifica – RP116154C2EF9AC8" and "Progetto di Ricerca RM11715C1743EE89"). The authors are also grateful to Prof. Fabio Altieri (Department of Biochemical Sciences, Sapienza University of Rome) for his helpful comments and Prof. Giuseppe Familiari (Section of Human Anatomy, Electron Microscopy Laboratory, Sapienza University of Rome,) for the SEM facilities provision and expertise.